\documentclass[11pt,twoside]{article}
\usepackage{preprint}
\usepackage{epsf}
\usepackage{psfig}
\def\FUSE{{\it FUSE}}
\def\arcsec{\ifmmode '' \else $''$\fi}
\def\arcmin{\ifmmode ' \else $'$\fi}
\def\arcsecpoint{\ifmmode ''\!. \else $''\!.$\fi}
\def\arcminpoint{\ifmmode '\!. \else $'\!.$\fi}
\def\cc{\ifmmode {\rm cm}^{-3} \else cm$^{-3}$\fi}
\def\cl{\ifmmode {\rm cm}^{-2} \else cm$^{-2}$\fi}
\def\micron{\ifmmode \mu{\rm m} \else $\mu$m\fi}
\def\kms{\ifmmode {\rm km\,s}^{-1} \else km\,s$^{-1}$\fi}
\def\Hubble{\ifmmode {\rm km\,s}^{-1}\,{\rm Mpc}^{-1}
        \else km\,s$^{-1}$\,Mpc$^{-1}$\fi}
\def\ergsec{\ifmmode {\rm ergs\;s}^{-1} \else ergs s$^{-1}$\fi}
\def\ergscm{\ifmmode {\rm ergs\,s}^{-1}\,{\rm cm}^{-2}
          \else ergs\,s$^{-1}$\,cm$^{-2}$\fi}
\def\ergscmA{\ifmmode {\rm ergs\,s}^{-1}\,{\rm cm}^{-2}\,{\rm \AA}^{-1}
          \else ergs\,s$^{-1}$\,cm$^{-2}$\,\AA$^{-1}$\fi}
\def\ergscmHz{\ifmmode {\rm ergs\,s}^{-1}\,{\rm cm}^{-2}\,{\rm Hz}^{-1}
          \else ergs\,s$^{-1}$\,cm$^{-2}$\,Hz$^{-1}$\fi}
\def\Msun{\ifmmode M_{\odot} \else $M_{\odot}$\fi}
\def\Lsun{\ifmmode L_{\odot} \else $L_{\odot}$\fi}
\def\qo{\ifmmode q_{0} \else $q_{0}$\fi}
\def\Ho{\ifmmode H_{0} \else $H_{0}$\fi}
\newcommand{\ovi}{O~{\sc vi}}

\begin{document}
\title{UV and X-ray Observations of Mass Outflows in AGN}
 \author{Gerard A. Kriss}
\affil{Space Telescope Science Institute}

\begin{abstract}
More than half of all low-redshift AGN exhibit UV and X-ray absorption by
highly ionized gas. The observed UV and X-ray absorption lines are almost
always blue-shifted at velocities of hundreds of km/s, indicating that the
absorbing gas is outflowing from the active nucleus. In some cases the
inferred mass flux rivals the Eddington limit of the central black hole,
an indication that these outflows are intimately related to the mass
accretion and energy generation mechanism in AGN. The ejected material
can also have an affect on the interstellar medium of the host galaxy
and the surrounding intergalactic medium. Over the past several years,
coordinated UV and X-ray observations of several bright AGN at high spectral
resolution using {\it HST}, {\it FUSE}, {\it Chandra}, and {\it XMM-Newton}
have contributed greatly to our understanding of these outflows.
I will give an overview of these recent observations,
summarize our {\it FUSE} survey of low-redshift AGN,
and interpret the results in the context of models of winds from accretion
disks and thermally driven winds from the obscuring torus.
\end{abstract}

\section{Introduction}

Mass outflows from active galactic nuclei (AGN) can profoundly affect the
evolution of the central engine \citep{BB99}, the host galaxy
and its interstellar medium \citep{SR98, WL03}
and also the surrounding intergalactic medium (IGM) \citep{Cavaliere02,
Granato04, SO04}.  Winds from the high
metal-abundance nuclear regions may be a significant source for
enriching the IGM \citep{Adelberger03}.
Absorption by the outflow can also collimate the ionizing radiation
\citep{Kriss97} and thereby influence the ionization
structure of the host galaxy and the surrounding IGM.

The outflowing gas in AGN is sometimes visible as extended, bi-conical emission
at visible or X-ray wavelengths (e.g., NGC 4151, \citealt{Evans93},
\citealt{Hutchings98}; NGC 1068, \citealt{Ogle03}),
but it most frequently manifests itself as blue-shifted
absorption features in their UV and X-ray spectra.
About half of all low-redshift AGN show X-ray absorption
by highly ionized gas \citep{Reynolds97, George98}, and a similar
fraction show associated UV absorption in ionized species such
as {\sc C~iv} \citep{Crenshaw99} and {\sc O~vi} \citep{Kriss01}.
In more luminous quasars, the fraction of objects in the Sloan Digital Sky
Survey that shows broad {\sc C~iv} absorption troughs rises steeply as the
troughs become narrower \citep{Tolea02,Reichard03}, comprising over 30\% of
the quasar population at widths narrower than 1000 \kms.
For AGN that have been observed in both the X-ray and the UV,
there is a one-to-one correspondence between
objects showing X-ray and UV absorption, implying that the phenomena are
related in some way (Crenshaw et al. 1999).
The high frequency of occurrence of UV and X-ray absorption suggests that the
absorbing gas has a high covering fraction, and that it is present in all AGN.
The gas has a total mass exceeding $\sim 10^3~\rm M_\odot$ (greater than the
broad-line region, or BLR), and is outflowing at a rate
$>0.1~\rm M_\odot~yr^{-1}$
($10 \times$ the accretion rate in some objects) \citep{Reynolds97}.

Key questions for understanding the outflowing, absorbing gas in AGN are:

\vspace{-12pt}
\begin{list}{} 
{\listparindent 0pt \leftmargin 0pt \rightmargin 0pt
\itemindent -0pt\parsep 0pt\labelsep 0pt\itemsep 0pt}
\begin{sloppy}

\smallskip
\item {$\bullet~~~$}
What are the column densities and ionic abundances in the absorbing gas?\\
Even such a basic question is uncertain since the absorption line profiles
are complex. Doublet ratios show that the absorbers can be optically thick,
but they are not black at line center. Thus column densities are frequently
underestimated, sometimes by as much as an order of magnitude
\citep{Arav02, Arav03}.

\smallskip
\item {$\bullet~~~$}
Where is the absorbing gas located?\\
The location of the absorbers is a vital clue to the process producing the
outflow. Winds arising from an accretion disk \citep{KK94, Murray95, Elvis00,
Proga00} will have
material at a broad range of radii reaching from near the disk itself to
beyond the BLR. Thermally driven winds arising from the
obscuring torus \citep{KK95, KK01} will lie at much larger radii,
typically at distances of $\sim 1$ pc.
The radial location also determines the mass flux---the larger the distance,
the higher the total mass and the mass flux in the wind.

\smallskip
\item {$\bullet~~~$}
Do variations reflect an ionization response, or are they due to bulk motion?\\
In the limited number of monitoring campaigns carried out thus far, examples
of both have been seen. The neutral hydrogen and the {\sc C~iii} absorption
in NGC~4151 responds quite clearly to continuum variations \citep{Kriss97,
Espey98}. Some absorbing clouds in NGC~3783 have shown changes
consistent with an ionization response, while others appear to be due to
bulk motion \citep{Crenshaw99, Gabel03b}.

\smallskip
\item {$\bullet~~~$}
How are the X-ray and UV absorption related?\\
In some cases, UV absorbing gas may be directly associated with the X-ray
warm absorber
(3C351: \citealt{Mathur94}; NGC~5548: \citealt{Mathur95};
NGC 3516: \citealt{Kraemer02}).
In the extensive recent {\it Chandra/FUSE/HST} campaign on NGC 3783
\citep{Kaspi02, Gabel03a}, the kinematics of the X-ray absorbing gas
are also a good match to the UV-absorbing gas.
In other cases, however, the UV gas appears to be in an even
lower ionization state, and there is no direct relation between the
X-ray absorption and the multiple kinematic components seen in the UV
(NGC~4151: \citealt{Kriss95}, \citealt{Kraemer01};
NGC~3516: \citealt{Kriss96a, Kriss96b};
NGC~5548: \citealt{Mathur99}, \citealt{Crenshaw03};
Mrk~509: \citealt{Kriss00b}, \citealt{Yaqoob03};
NGC~7469: \citealt{Kriss00a, Kriss03, Blustin03}).
The X-ray absorbing gas itself contains material spanning a large range
of ionization parameters \citep{Lee02, Sako03, Netzer03},
and it is likely that this broad range of physical
conditions can also include the UV-absorbing ions. This is a natural prediction
of the thermally driven wind model of \citet{KK95, KK01}, and would also be
likely in disk-driven winds.

\end{sloppy}
\end{list}

\vspace{-8pt}
To address some of these questions, we have been conducting a survey of the
$\sim100$ brightest AGN using the 
{\it Far Ultraviolet Spectroscopic Explorer (FUSE)}.
The short wavelength response (912--1187 \AA) of {\it FUSE} \citep{Moos00}
enables us to make high-resolution spectral measurements
($R \sim 20,000$) of the high-ionization
ion {\sc O~vi} and the high-order Lyman lines of neutral hydrogen.
As of November 1, 2002, we have observed a total of 87 AGN.
Of these, 57 have $z < 0.15$, so that the {\sc O~vi} doublet is visible
in the {\it FUSE} band.
The \ovi\ doublet is a crucial link between
the higher ionization absorption edges seen in the X-ray and the lower
ionization absorption lines seen in earlier UV observations.
The high-order Lyman lines provide a better constraint on the
total neutral hydrogen column density than Ly$\alpha$ alone.
Lower ionization species such as {\sc C~iii} and {\sc N~iii} also have strong
resonance lines in the \FUSE\ band, and these often are useful for setting
constraints on the ionization level of any detected absorption.
The Lyman and Werner bands of molecular hydrogen also fall in the \FUSE\ band,
and we have searched for (but not found) intrinsic $\rm H_2$ absorption that
may be associated with the obscuring torus.

\begin{figure}[h]
\plottwo{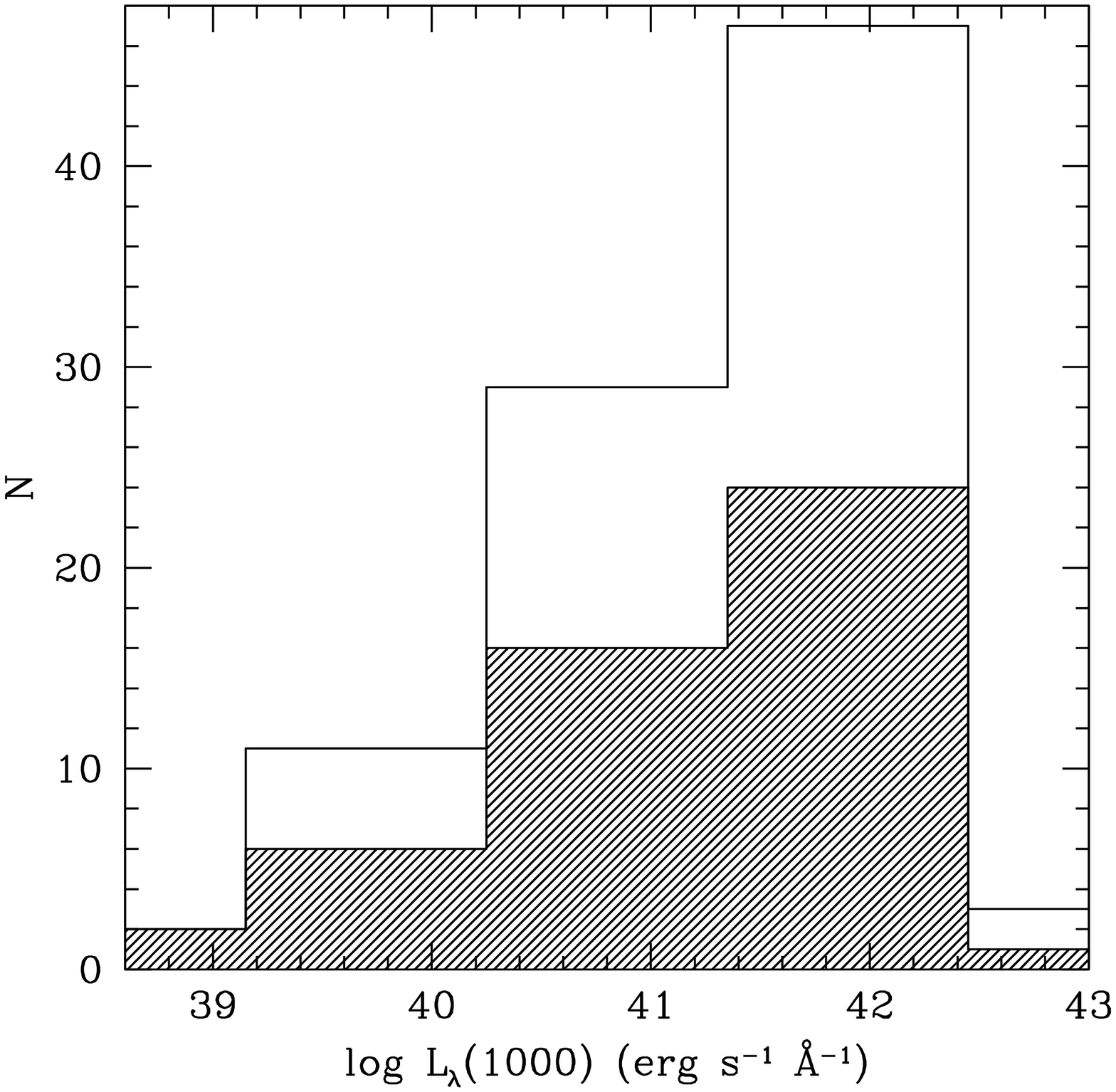}{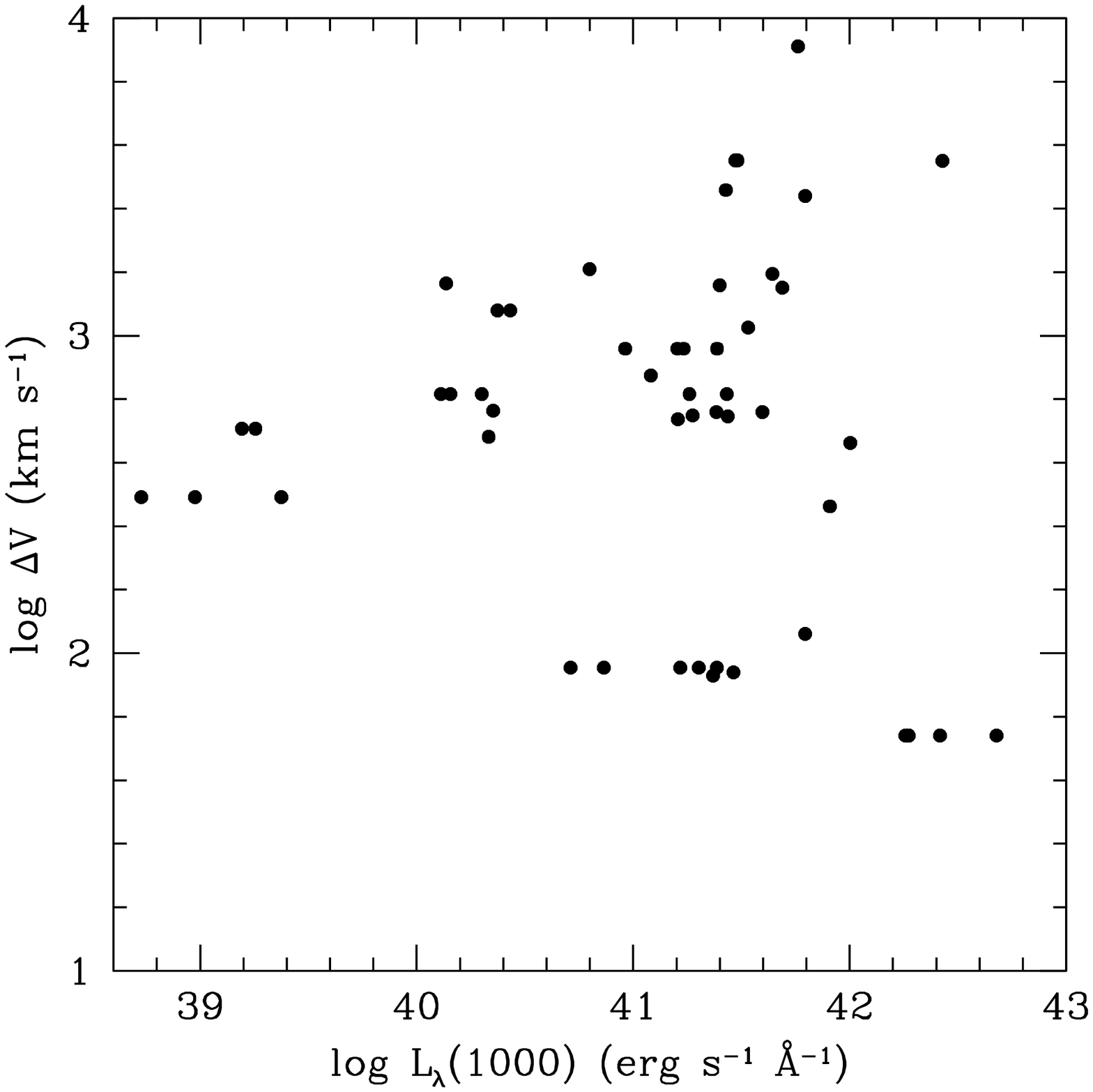}
\caption{
Left: Histogram of FUSE AGN versus luminosity. The shaded area shows the number of objects exhibiting intrinsic absorption. Right: The points show outflow
velocity as a function of luminosity.
\label{fig1}}
\end{figure}

\section{Survey Results}

As shown in the left panel of Figure 1, absorption is common at all
luminosities, and over 50\% (30 of 53) of the low-redshift Type 1 AGN
observed using \FUSE\ show detectable \ovi\ absorption,
comparable to those Seyferts that show
longer-wavelength UV \citep{Crenshaw99}
or X-ray \citep{Reynolds97, George98} absorption.
None show intrinsic $\rm H_2$ absorption.
We see three basic morphologies for \ovi\ absorption lines:
(1) {\bf Single}: 13 of 30 objects exhibit single, narrow,
isolated \ovi\ absorption lines, as
illustrated by the spectrum of Ton~S180 \citep{Turner01}.
PG0804+761, shown in the top panel of Figure 2, is another example.
(2) {\bf Blend}: multiple \ovi\ absorption components that are blended
together.
10 of 30 objects fall in this class, and the spectrum of Mrk~279
is typical \citep{Scott04}.
The middle panel of Figure 2 shows Mrk 478 as another example.
(3) {\bf Smooth}: The 7 objects here are an extreme expression of the
``blend" class,
where the \ovi\ absorption is so broad and blended that individual
\ovi\ components cannot be identified.  NGC~4151 typifies
this class \citep{Kriss92, Kriss95, Kriss01}.
The mini-BAL QSO PG1411+442 is another example of this class, shown in the
bottom panel of Figure 2.

Individual \ovi\ absorption components in our spectra have
FWHM of 50--750 \kms, with most objects having FWHM $< 100~\kms$.
The multiple components that are typically present are almost always
blue shifted, and they span a velocity range of 200--4000 \kms;
half the objects span a range of $< 1000~\kms$.
As shown in the right panel of Figure 1, the maximum outflow velocities show
a tendency to increase with source luminosity, perhaps indicating that
radiative acceleration plays some role in the dynamics.
Note also that there is a population of low-velocity absorbers present
at all luminosities.

\begin{figure}[ht]
\plotfiddle{"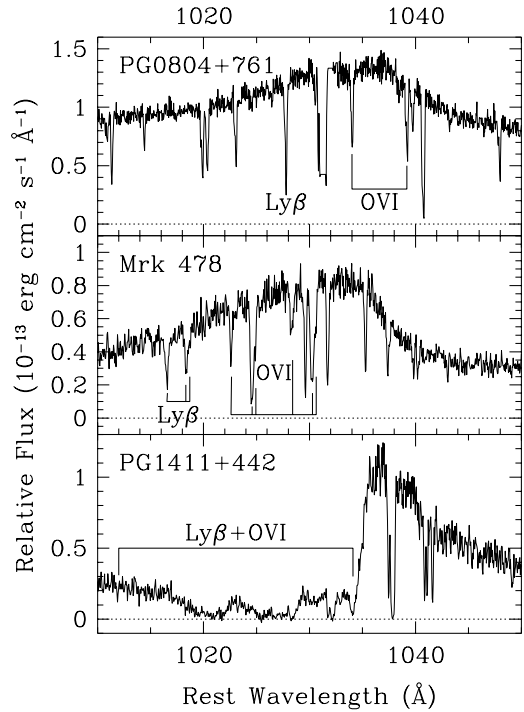"}{3.4in}{0}{39}{39}{-130}{-30}
\caption{
The three classes of O VI absorption line morphology: Single, isolated lines: PG0804+761 (top); Multiple, blended lines: Mrk 478 (middle); Broad, blended
trough: PG1411+442 (bottom).
\label{fig2}}
\end{figure}

\section{Discussion}

The multiple kinematic components frequently seen in the UV absorption spectra
of AGN clearly show that the absorbing medium is complex, with separate
UV and X-ray dominant zones.
In some cases, the UV absorption component corresponding to the X-ray warm
absorber can be clearly identified (e.g., Mrk~509, \citealt{Kriss00b}).
In others, however, {\it no} UV absorption component shows physical
conditions characteristic of those seen in the X-ray absorber
(NGC~3516, \citealt{Kriss96a, Kriss96b}; NGC~5548, \citealt{Brotherton02}).
One potential geometry for this complex absorbing structure is high-density,
low-column UV-absorbing clouds embedded in a low-density,
high-ionization medium that dominates the X-ray absorption.

Disk-driven winds are a possible explanation for some cases of AGN outflows.
By analogy to stellar winds, one would expect the terminal velocity of an
AGN outflow to reflect the gravity of its origin.
Disk-driven winds should therefore have velocities in the range of several
thousand \kms.
Objects with broad, smooth profiles might fall in this category.
The geometry proposed by \citet{Elvis00} suggests that these objects should
have only modest inclinations. However, two prime examples of Seyferts with
broad smooth absorption troughs, NGC 3516 \citep{Hutchings01} and
NGC 4151 \citep{Kriss01}, are likely the
highest inclination sources in our sample given their extended, bi-conical
narrow emission-line region morphologies \citep{Miyaji92, Evans93} and their
opaque Lyman limits \citep{Kriss97}.

The lower velocities we observe in objects like NGC 3783 and NGC 5548
are more compatible with thermally driven winds
from the obscuring torus \citep{KK95, KK01}.
In these thermally driven winds, photoionized evaporation in the presence of
a copious mass source (the torus) locks the ratio of ionizing intensity to
total gas pressure (the ionization parameter $\Xi$) at a critical value.
For AGN spectral energy distributions lacking a strong extreme ultraviolet
bump, such as the composite spectra of quasars assembled by \citet{Zheng97},
\citet{Laor97}, and \citet{Telfer02},
the ionization equilibrium curve exhibits an extensive vertical branch.
Thus, at the critical ionization parameter for evaporation, there is a broad
range of temperatures that can coexist in equilibrium at nearly constant
pressure.
For this reason, the flow is expected to be strongly inhomogeneous.
Outflow velocities are typical of the sound speed in the heated gas, or
several hundred \kms, comparable to the velocities seen in many AGN.

In summary, we find that \ovi\ absorption is common in low-redshift ($z < 0.15$)
AGN. 30 of 53 Type 1 AGN with $z < 0.15$ observed using \FUSE\ show
multiple, blended \ovi\ absorption lines with typical widths
of $\sim 100~\kms$ that are blueshifted over a velocity range of $\sim$
1000 \kms.
Those galaxies in our sample with existing X-ray or longer wavelength UV
observations also show {\sc C~iv} absorption and evidence of a soft X-ray
warm absorber.
In some cases, a UV absorption component has physical properties
similar to the X-ray absorbing gas, but in others there is no clear
physical correspondence between the UV and X-ray absorbing components.

\acknowledgments{
I thank all the members of the FUSE AGN Working Group for their contributions
to this research.
This work is based on data obtained for the Guaranteed Time Team by the
NASA-CNES-CSA FUSE mission operated by the Johns Hopkins University. Financial
support to U. S. participants has been provided by NASA contract NAS5-32985.
}

\end{document}